\documentstyle[aps,prl,twocolumn]{revtex}
\input epsf
\begin{document}
\title{Half-Periodic Josephson Effect in an $s$-Wave Superconductor - 
Normal Metal -$d$-Wave Superconductor Junction}
\author{Alexandre M. Zagoskin}
\address{Department of Physics and Astronomy, The University of British
Columbia, 6224 Agricultural Rd., Vancouver, B.C., V6T 1Z1, Canada; Email
zagoskin@physics.ubc.ca} 
\date{Preprint: cond-mat@9702123}
\maketitle

\begin{abstract}
We predict that the Josephson current in a clean 
$s$-wave superconductor-normal metal-$d$-wave superconductor junction is periodic in superconducting phase
difference $\varphi$ with period $\pi$ instead of $2\pi$. The frequency of
non-stationary  Josephson effect is correspondingly $2\omega_J = 4eV.$
The effect is due to coexistence in the normal layer
of current carrying Andreev levels with
phase differences $\varphi$ and $\varphi+\pi.$
\end{abstract}

\begin{figure}
\epsfysize=4 in
\epsfbox{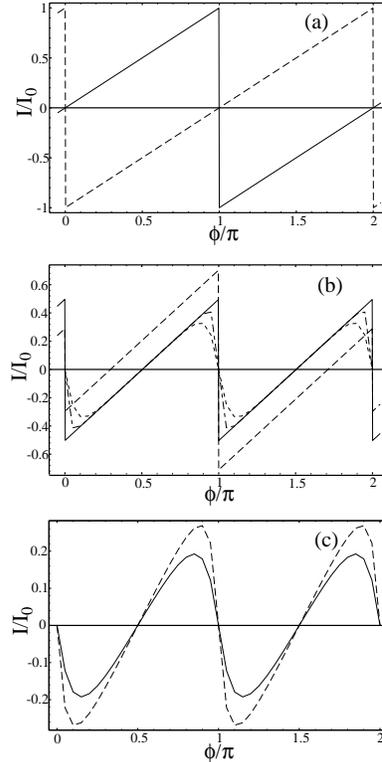}
\caption{(a) Current-phase characteristics of a clean SNS junction at
zero temperature: conventional ("zero"-) junction (solid line); $\pi$-junction (dashed line).
(b) Current-phase characteristics of half-periodic Josephson current: 
$Z=0, T=0$ (solid line); $Z=2^{1/2}-1, T=0$ (dashed line); $Z=0, L/l_T=0.2$
(dash-dotted line); $Z=0, L/l_T=0.5$ (dotted line). Disbalance function $Z$ 
is defined in Eq.(9).
(c) Current-phase characteristics of half-periodic Josephson current
at finite normal scattering: $Z=0; L/l_{i}=0.05$ (dashed line); $0.1$
 (solid line).}
\label{f.2}
\end{figure}

The Josephson effect in superconductor- normal metal -superconductor (SNS)
junctions differs from the effect in tunneling junctions in several important aspects \cite{Barone2}. In particular, in a long clean SNS junction at $T=0$ the Josephson current, $I_J(\varphi)$, is a sawtooth function of superconducting phase
difference $\varphi$\cite{Ishii,Bratus,Bardeen}
 (solid line in Fig.\ref{f.2}a), in contrast to 
$I_c\sin\varphi$ dependence in tunneling 
  junctions\cite{Barone2}. The reason of this behaviour is that the Josephson
current is transferred  through the normal layer by current carrying
Andreev levels, formed  due to subgap Andreev reflections 
of electrons
and holes  from the non-diagonal pairing potential of superconductors
\cite{Bardeen,Kulik}. Positions of these levels, and therefore the current, depend on  $\varphi$.

\begin{figure}
\epsfysize=3 in
\epsfbox{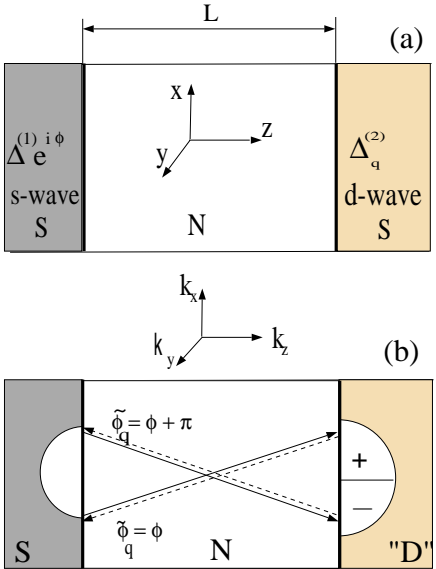}
\caption{(a) $s$-Wave superconductor - normal metal - $d$-wave superconductor
junction ("SND"-junction). We can choose real order parameter in $d$-wave superconductor.
 (b) Zero- and $\pi$- Andreev levels in a clean SND junction. }
\label{f.1}
\end{figure}

In this Letter we consider the case when one of the superconductors
has $d$-wave pairing symmetry:  $s$-wave superconductor - 
normal metal -$d$-wave superconductor ("SND") junction (Fig.\ref{f.1}a).
There is a strong evidence that such pairing is realized in superconducting cuprates, based particularly on phase sensitive measurements \cite{AMURRU,Tsuei}.
The experimental realization of SND junction is feasible, and could
provide additional information about superconductivity in cuprates.
Both tunneling and SNS junctions between conventional and $d$-wave superconductors are now being actively investigated\cite{SID}.

We predict that SND junctions will
combine properties of conventional and 
   so called $\pi$-junction\cite{Barone,Sigrist}  with negative
critical current\cite{footnote2}(Fig.\ref{f.2}a, dashed line).
 As a result, they exhibit {\em half-periodic Josephson effect}, with
$I(\phi)$ being $\pi$-periodic function of phase (instead of $2\pi$),
which can be detected, e.g., by doubling of frequency of non-stationary 
Josephson effect.

The physical picture  of the effect is as follows.
In $d$-wave superconductors the order parameter can have either sign depending
on the momentum direction on the Fermi surface, which can be treated 
as an additional, intrinsic phase $\pi$\cite{Sigrist-Ueda}. 
\noindent
Therefore
the Andreev levels in SND junction depend (instead of $\varphi$) on effective phase difference  \begin{equation}
\tilde{\varphi}^{({\bf k}_{\parallel})} = \varphi + \pi\:\: \vartheta(-\Delta_{\hat{q}}),
\label{3}
\end{equation}
 where $\vartheta(x)$ is the Heaviside step function. 
Unit vector $\hat{q}$ gives the direction of the
wave vector of transmitted state    in  $d$-wave 
superconductor, ${\bf q}$, with components ${\bf k}_{\parallel}$ (parallel to the interface) and $q_z(E)$
(Fig.\ref{f.1}b). There are thus two sets of Andreev levels in the normal layer:
with $\tilde{\varphi} = \varphi$ ($\Delta_{\hat{q}}$ positive) and
$\tilde{\varphi} = \varphi+\pi$ ($\Delta_{\hat{q}}$ negative).

\noindent
The Josephson current through   
$\pi$-levels 
will be a $2\pi$-periodic function of phase,  shifted
   by $\pi$ with respect to the current 
carried by zero levels.
  The resulting current  is $\pi$-periodic in phase 
(Fig.\ref{f.2}b).

From the Josephson relation $\partial_t\varphi = 2eV$
  we see that the Josephson frequency   
doubles: $\tilde{\omega}_J = 2\omega_J=4eV$.

In our further analysis we assume that the following conditions are satisfied:
\begin{equation}
\max(\xi_0) \ll L \ll l_T,l_\varepsilon,l_i. \label{0}
\end{equation}
where $L$ is the width of the normal metal layer, 
$\max(\xi_0)$ the larger of the correlation length in
either superconductor,  $l_i, l_{\varepsilon}$ are elastic and inelastic scattering length respectively,   and $l_T =  v_F/ (2\pi k_B T)$ 
is  the   normal metal coherence length in the clean limit.  
The latter is the characteristic coupling length in SNS 
junction\cite{Ishii,Bratus,Kulik},   and can be much larger than
 $\xi_0$ (especially in case of cuprates  \cite{YBCO}).
In silver, e.g., at $1 K$ $l_T = 1.67 \cdot 10^{-4}$ cm, compared to
$\xi_0 \sim 10^{-5}$ cm typical for conventional superconductors.

Andreev levels in the normal layer are obtained
by solving  Bogoliubov-de Gennes equations 
for the two-component wave function in both
superconductors and normal layer, separately for each 
 ${\bf k}_{\parallel}$-mode, and matching the wave functions at the interfaces
\cite{BdG}; for the SND junction we have: 
\begin{eqnarray}
\left(\begin{array}{ll}
{\cal H}-\frac{k_F^2-k_{\parallel}^2}{2m} & \Delta(z)  \\
\Delta^*(z)  & -\left({\cal H}-\frac{k_F^2-k_{\parallel}^2}{2m}
\right)
\end{array}\right)\left(\begin{array}{l} u(z) \\ v(z)
\end{array}\right) = E \left(\begin{array}{l} u(z) \\ v(z)
\end{array}\right). \label{1}
\end{eqnarray}
 Here ${\cal H} = -\frac{1}{2m}\nabla_z^2$ is the one-particle 
Hamiltonian; $\Delta(z)$ is the non-diagonal
potential (which is zero in normal layer, $\Delta^{(1)}e^{i\varphi}$
 in  $s$-wave superconductor, and $\Delta^{(2)}_{\hat{q}}$ in  $d$-wave superconductor; we choose real $\Delta^{(1)},\Delta^{(2)}_{\hat{q}}$).  
This standard approximation is justified by the condition $\xi_0\ll L$
\cite{BdG}. We assume for simplicity the same value of $k_F$ in all three 
regions. The corrections due to differences in $k_F$ and finite
normal scattering on the interface will be considered elsewhere.

In the normal layer the normal component of the momentum of an electron (hole)
with energy $E$ and  tangential momentum $ {\bf k}_{\parallel}$
is  $k_z^{e,h}(E) = \sqrt{k_F^2-k_{\parallel}^2 \pm 2mE}$.
 In the superconductor, for a subgap 
quasiparticle, $E<|\Delta|$, it  transforms into
\begin{eqnarray}
q_z^{\pm}(E) =  k_z(0)\sqrt{1 \pm  i \frac{|\Delta|}{k_z(0)^2/2m}
\sqrt{1-\frac{E^2}{|\Delta|^2}}}.
   \label{4}
\end{eqnarray}
If $|\Delta| \ll \frac{k_z^2(0)}{2m}$, then $\Re q_z^{\pm}\approx k_z(0)$,
 i.e. the quasiparticle  momentum does not change direction in 
the superconductor: $\hat{q} \approx \hat{k}$, 
${\bf k}=({\bf k}_{\parallel},k_z).$  Therefore   the momentum of the quasiparticle
in the normal layer  determines the value $\Delta_{\hat{k}}$
 of the $d$-wave order parameter entering  Bogoliubov-de Gennes equations
(\ref{1}), and   the effective phase difference
for given Andreev level (Fig.\ref{f.1}b). 

Under the above conditions the energies of low-lying Andreev levels
($E\ll|\Delta|$) are given by
\begin{equation}
\left(k_z^e(E) 
- k_z^h(E)\right)L \pm  \tilde{\varphi}^{({\bf k}_{\parallel})}
 = \pi (2 n + 1); n=0,\pm1,\pm2,\dots. \label{2}
\end{equation}
This is a direct generalization of the result by Kulik\cite{Kulik}.

The Josephson current through the contact can be calculated 
using the low-energy excitation spectrum determined from (\ref{2})\cite{Bardeen,WanWees}, to yield at
$T=0$:
\begin{equation}
I_J(\varphi) =
\sum_{\vec{\kappa}}  \frac{2ev_{Fz}^{(\vec{\kappa})}}
{\pi L} 
\sum_{n=1}^{\infty}
(-1)^{n+1}\frac{\sin n\tilde{\varphi}^{(\vec{\kappa})}}{n}. \label{5}
\end{equation}
Here $\tilde{\varphi}^{(\vec{\kappa})}$ is the effective phase shift,
and $v_{Fz}^{(\vec{\kappa})}$ is the normal to the interface component
of Fermi velocity in a mode with tangential momentum ${\bf k}_{\parallel}
=\vec{\kappa}$:
$\left(mv_{Fz}^{(\vec{\kappa})}\right)^2 + \vec{\kappa}^2 =
k_F^2.$ The summation $\sum_{\vec{\kappa}}  = 
S \int_{|\vec{\kappa}|\leq k_F} d\vec{\kappa}_x\:d\vec{\kappa}_y$   
is extended over all allowed tangential modes\cite{footnote}; $S$ is the area of the junction.

It follows from (\ref{5}) that  
the  Josephson current in clean  SND junction
is  indeed  a sum of independent contributions of zero- and $\pi$-modes:
\begin{eqnarray}
I_J(\varphi) = I_0\left( \alpha^{(+)}F(\varphi)+\alpha^{(-)}F(\varphi\!\!+\!\!\pi)
\right)=
\label{6}\\
= \frac{I_0}{2}
\left(F(\varphi)\!\!+\!\!F(\varphi\!\!+\!\!\pi)+Z \left(F(\varphi)\!\!-\!\!
F(\varphi\!\!+\!\!\pi)\right)
\right).\label{7}
\end{eqnarray}

\begin{figure}
\epsfysize=4 in
\epsfbox{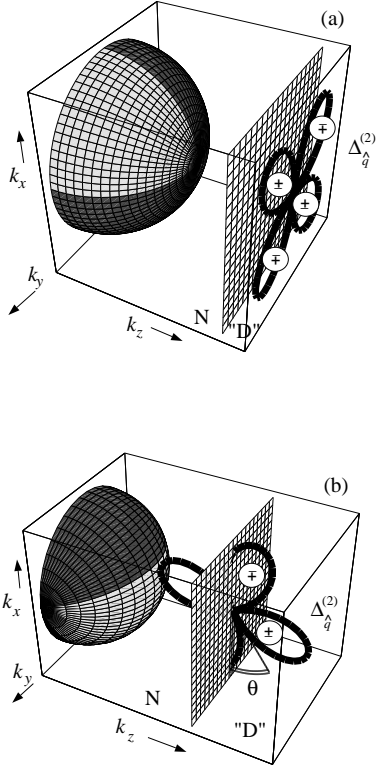}
\caption{On calculation of disbalance function at different orientations of $d$-wave
 superconductor with respect to the interface: white regions on the Fermi 
surface (in normal metal) correspond to zero modes, shadowed - to $\pi$-modes,
or vice versa.
(a) Interface parallel to $x^2-y^2$-plane. (b) Interface normal to 
$x^2-y^2$-plane; $\theta$ is the  angle between the SN interface 
and the nodal direction in $x^2$-$y^2$-plane.}
\label{f.3}
\end{figure}

\noindent
Here $F(x) = \frac{2}{\pi}\sum_{n=1}^{\infty}
(-1)^{n+1}\frac{\sin nx}{n}$ is the $2\pi$-periodic sawtooth of unit amplitude;
$I_0 = \frac{e}{L}\sum_{\vec{\kappa}}  v_{Fz}^{(\vec{\kappa})}$ is the critical current of clean SNS junction
at zero temperature \cite{Bratus,Bardeen}; and weight factors $\alpha^{(\pm)} = \sum_{\vec{\kappa}} ^{\pm} v_{Fz}^{(\vec{\kappa})}/\sum_{\vec{\kappa}}  v_{Fz}^{(\vec{\kappa})}$, where the $\pm$-summation is extended over 
zero- and $\pi$-modes respectively. 

The ratio between the two contributions in (\ref{6}) generally 
depends on the orientation of the $d$-superconductor  with respect to the interface, parametrized by some angles $\theta_a$, through the disbalance function,
\begin{equation}
Z(\theta_a) = \frac{\alpha^{(+)}(\theta_a)-\alpha^{(-)}(\theta_a)}
{\alpha^{(+)}(\theta_a)+\alpha^{(-)}(\theta_a)}; \:\: |Z(\theta_a)| \leq 1;
\end{equation} 
$Z=\pm 1$ corresponds to  purely zero- or $\pi$- junction\cite{Barone,Sigrist} respectively;  
$Z=0$  when amplitudes of the contributions from
zero- and $\pi$-levels are equal.

 Due to nonsinusoidal form of the current-phase characteristics, the current is not zero  if   $Z(\theta_a)=0$
(as it would be in case of tunneling contact \cite{Sigrist},
with $\sin \varphi$ in place of $F(\varphi)$\cite{Tanaka}). On the contrary, 
the resulting current  is $\pi$-periodic in phase 
(Fig.\ref{f.2}b), and its critical value is halved.

The simplest case is an SND junction where $x^2\!\!-\!\!y^2$-plane of  
the $d$-wave superconductor is parallel to the interface (Fig.\ref{f.3}a):
then $Z(\theta_a)\equiv 0$ by symmetry. For cuprates, this (001)-plane 
is also the easiest cleavage plane, which makes it the best candidate for 
experimental observation of the effect.  

Another possibility is to choose the interface normal to that plane.
 It is easy to see that in this case the  disbalance 
function depends on a single angle $\theta$ 
between the SN interface 
and the nodal direction in $x^2$-$y^2$-plane
(see Fig.\ref{f.3}b), and
$Z(\theta) =   
\pm \left(\sqrt{2}\cos[\theta-\frac{\pi}{4}] - 1\right); \:\:\: 
(0\leq\theta\leq\frac{\pi}{2}).$
 Since $|Z(\theta)| \leq (\sqrt{2}-1)$,
the Josephson current in this case  always contains a $\pi$-periodic component (see Fig.\ref{f.2}b).

It is worth noting that in 
case of "DND" junction between two $d$-wave
superconductors the half-periodic Josephson effect is present as well, 
but it  disappears at
certain orientations of the crystals.
 If the junction interface is parallel
to  $x^2\!\!-\!\!y^2$-plane of both superconductors, then $
Z(\alpha) =   \pm \left( 1 - 4\frac{\alpha}{\pi}\right).$ Here
$0\leq\alpha\leq\frac{\pi}{2}$ is the angle between nodal directions 
in the superconductors, and the effect is absent if 
$\alpha = 0$ or $\frac{\pi}{2}.$
  
So far we did not consider the effects of nonmagnetic impurity scattering
and finite temperature.

At finite temperatures the condition $E \ll |\Delta^{(i)}|$ will be
 met for almost all
 of the Fermi surface if 
\begin{equation}k_B T \sim E \ll |\Delta^{(1)}|; {\rm max}\: |\Delta^{(2)}_{\hat{k}}|;
\end{equation}
 corrections to the current from the regions close to the nodal lines will be
small by the same parameter,  $k_BT/{\rm max}\: |\Delta^{(2)}_{\hat{k}}|$. This   additional  condition on temperature  is in fact less restrictive that the one following from (\ref{0}).

 Weak  scattering   can be taken  into account   via broadening of
Andreev levels due to finite elastic scattering time, $\tau = l_i/v_F$ \cite{Kulik-Mitsai}.
Using the approach of \cite{kadig}, we find for $L/l_i,L/l_T \ll 1$:
\begin{eqnarray}
I_J(\varphi) =
\sum_{\vec{\kappa}}  \frac{2ev_{Fz}^{(\vec{\kappa})}}
{\pi L} 
\sum_{n=1}^{\infty}
(-1)^{n+1} \exp\left[-2n\frac{L}{l_i^{(\vec{\kappa})}}\right]
\times\nonumber\\
\times
\frac{(L/l_T^{(\vec{\kappa})}) \sin n\tilde{\varphi}^{(\vec{\kappa})}}{\sinh (nL/l_T^{(\vec{\kappa})})} \label{11}
\end{eqnarray}
(see Fig.\ref{f.2}b,c). Here $l_i^{(\vec{\kappa})} = l_i\frac{v_{Fz}^{(\vec{\kappa})}}{v_F},
l_T^{(\vec{\kappa})} = l_T\frac{v_{Fz}^{(\vec{\kappa})}}{v_F}$ 
are effective scattering and coherence length in mode 
$\vec{\kappa}$ respectively. The effect survives finite temperature
 and  weak elastic scattering in the normal layer. 

In the limit of strong elastic scattering, $L \gg l_i$, or at higher temperatures, $L \gg l_T$,   the sawtooth current-phase
dependence in long SNS junctions reduces to the sinusoidal one
\cite{Kulik-Mitsai}, and  the effect disappears   due to cancellation of contributions from zero- and $\pi$-modes. 

{\em Note added:}\\
\noindent
After this paper was submitted, the preprint \cite{Bagwell2} appeared, where 
analogous conclusions were made in the case of short SND junction ($L\ll
\min(\xi_0)$).

 \acknowledgements{I would like to thank I.~Affleck, D.~Bonn, 
A.~Dubin, D.L.~Feder, I.~Herbut, S.~Kivelson,
M.~Oshikawa, and P.C.E.~Stamp for fruitful discussions and criticism,
and  P.F.~Bagwell and Y.~Tanaka for kindly
acquainting me with the results of their current research.}

\end{document}